\documentclass[fleqn,usenatbib]{mnras}

\usepackage{newtxtext,newtxmath}
\usepackage{graphicx,graphics,color}
\usepackage{xcolor}
\usepackage[T1]{fontenc}
\usepackage[flushleft]{threeparttable}

\definecolor{byzantine}{rgb}{0.74, 0.2, 0.64}

\newcommand\kev{{\rm~keV}}

\newcommand\kms{\ifmmode {\rm~km\ s}^{-1} \else ~km s$^{-1}$\fi}
\newcommand\Hunit{\ifmmode {\rm~km\ s}^{-1}\ {\rm Mpc}^{-1}
        \else ~km s$^{-1}$ Mpc$^{-1}$\fi}
\newcommand\ctssec{\ifmmode {\rm~count\ s}^{-1} \else ~count s$^{-1}$\fi}
\newcommand\ergsec{\ifmmode {\rm~erg\ s}^{-1} \else
        ~erg s$^{-1}$\fi}
\newcommand\funit{\ifmmode {\rm~erg\ s}^{-1}\;{\rm cm}^{-2} \else
        ~ergs s$^{-1}$ cm$^{-2}$\fi}
\newcommand\phflux{\ifmmode {\rm~photon\ s}^{-1}\;{\rm cm}^{-2}
        \else   ~photon s$^{-1}$ cm$^{-2}$\fi}
\newcommand\efluxA{\ifmmode {\rm~erg\ s}^{-1}\;{\rm cm}^{-2}\;{\rm
        \AA}^{-1} \else ~erg s$^{-1}$ cm$^{-2}$ \AA$^{-1}$\fi}
\newcommand\efluxHz{\ifmmode {\rm~erg\ s}^{-1}\;{\rm cm}^{-2}\;{\rm
        Hz}^{-1} \else ~erg s$^{-1}$ cm$^{-2}$ Hz$^{-1}$\fi}
\newcommand\cc{\ifmmode {\rm~cm}^{-3} \else cm$^{-3}$\fi}
\newcommand\FWHM{\ifmmode {\rm~FWHM} \else ${\rm~FWHM}$\fi}
\newcommand\Zsun{\ifmmode Z_{\odot} \else $M_{\odot}$\fi}
\newcommand\Lsun{\ifmmode L_{\odot} \else $L_{\odot}$\fi}

\newcommand\hbeta{\ifmmode {\rm H}\beta \else H$\beta$\fi}
\newcommand\Kalpha{\ifmmode {\rm K}\alpha \else K$\alpha$\fi}
\newcommand\nh{\ifmmode N_{\rm H} \else N$_{\rm H}$\fi}

\newcommand{\lum}{erg\,s$^{-1}$}

\newcommand{\Msun}{\ensuremath{\rm M_{\odot}}}
\DeclareRobustCommand{\VAN}[3]{#2}
\let\VANthebibliography\thebibliography
\def\thebibliography{\DeclareRobustCommand{\VAN}[3]{##3}\VANthebibliography}

\usepackage{graphicx}	
\usepackage{amsmath}	

\title[Abell 725]{GMRT unveils steep-spectrum antique filaments in the galaxy cluster Abell~725}

\author[M. B. Pandge et al.]{
M. B. Pandge,$^{1}$\thanks{E-mail: mbpandge@gmail.com (DSCL)}
Ruta Kale,$^{2}$ Pratik Dabhade,$^{3,4}$ Mousumi Mahato$^{3}$ and 
\newauthor Somak Raychaudhury$^{3,5}$
\\
$^{1}$DST-INSPIRE Faculty, Dayanand Science College, Barshi Road, Latur 413512, Maharashtra, India\\
$^{2}$National Centre for Radio Astrophysics (TIFR), Pune 411007, Maharashtra, India\\
$^{3}$Observatoire de Paris, LERMA, Coll\`ege de France, CNRS, PSL University, Sorbonne University, 75014, Paris, France\\
$^{4}$Inter-University Centre for Astronomy and Astrophysics, Pune 411007, Maharashtra, India\\
$^{5}$School of Physics and Astronomy, University of Birmingham,Birmingham B15~2TT, UK\\
}
\date{Accepted XXX. Received YYY; in original form ZZZ}
\pubyear{2021}
\begin{document}
\label{firstpage}
\pagerange{\pageref{firstpage}--\pageref{lastpage}}
\maketitle

\begin{abstract}
We present original GMRT radio observations of the galaxy cluster Abell~725, at a redshift of  0.09, along with other archival observations. Our GMRT maps reveal two steep-spectrum diffuse  filaments in the cluster, along with a previously reported arc-like structure, and a wide-angle tail (WAT) radio source associated with the Brightest Cluster Galaxy (BCG) at the periphery of the cluster.
The bent morphology of the WAT indicates that its jets have been swept back by the dynamic pressure resulting from the motion of the BCG through the surrounding intracluster medium. The BCG associated with the WAT hosts a black hole whose mass we estimate to be 1.4$\pm0.4 \times10^{9} \Msun$.
We observe a 2\arcmin (195\,kpc in projection) offset between the BCG and the X-ray centroid of the galaxy cluster, which, along with other dynamic features, indicates the cluster's early stage of evolution. The WAT radio galaxy, the arc and the filaments have spectral indices $\alpha_{612}^{240}= -0.46\pm 0.15$,  $-0.8\pm0.3$, and ($-1.13\pm 0.48$, $-1.40\pm 0.50$), respectively. The WAT and the arc are connected structures, while the filaments are detached from them, but are found to be along the trail of the WAT. Based on the morphology of the components, and the progressive steepening of the components from the core of the WAT to the filaments, we propose that this system is a radio galaxy with trailing antique filaments.

\end{abstract}




\section{Introduction}
Powerful diffuse radio sources are mostly found in major merging galaxy clusters, indicating that their origin is likely connected to the process of merger \citep{cohen2011}. Clusters of galaxies often show extended synchrotron emission in the intra-cluster medium \citep{Feretti2005}. These radio sources mostly have very low surface brightness and steep radio spectrum, with their overall extent growing to megaparsec scales. Such radio sources, associated with clusters of galaxies, have been classified as radio halos and radio vestigials. The radio halos, with low polarisation, are found to be at the centres of the galaxy clusters, while radio relics, found towards the periphery of clusters, with elongated arc-like shapes, have emissions with higher levels of polarisation. The exact physics leading to the formation of these sources is indeed still a matter of debate, such sources are clearly associated with particle acceleration processes in the context of cluster mergers \citep[e.g.][]{vanWeeren19}. Apart from radio halos and relics, the only sources growing to Mpc scales are the giant radio galaxies (sizes $>$\,0.7~Mpc), which are in general not found in clusters \citep[e.g.][]{DabhadeSagan1}.

The narrow-angle tail (NAT) and wide-angle tail (WAT) radio galaxies are generally found in the galaxy cluster environment, where the radio jets and lobes, in these systems, often appear to be bent in a common direction \citep{Miley,1985AJ.....90..954O,1997ApJS..108...41O,2019Ap&SS.364...72P}. In cases where the angle $\phi$ between two tails of the radio galaxy is $\phi\leq 90\degr$, forming a `V' shape, it is referred to as NAT (e.g NGC 1262), while if the angle is $90\degr\leq \phi \leq 180\degr$, it is known as a WAT and is usually found in merging or not-relaxed clusters \cite{1990AJ.....99...14B}. The relative  velocities of these sources, with respect to the ICM of the host cluster, are thought to be sufficient for the tails to be bent, due to the action of ram pressure interactions \citep{1985AJ.....90..954O}. 


Duty cycles of radio galaxies are a probe of the accretion at the supermassive blackhole. In the cases of powerful double radio sources of FR II kind, two or more phases can be identified due to the well defined double structure formed by the hot-spots. However in FR I type of sources, the separation of different episodes of activity is difficult \citep{2007MNRAS.378..581J}. 
After the jets stop, the lobes evolve passively while interacting with the surrounding medium forming remnant lobes that may or may not show the connection to the host AGN. This phase is known to be short-lived due to the rapid loss of energy due to adiabatic expansion and synchrotron radiation. For FRI sources at the centres of massive galaxy clusters or groups, the episodes of the activity can be inferred from the detection of cavities in the X-ray emission from the intra-cluster medium (ICM) \citep{2001ApJ...558L..15B,2004ApJ...607..800B,2010ApJ...712..883D,2012MNRAS.421..808P}. 
The remnant radio galaxies in clusters may passively evolve and fade or could be rejuvenated 
by compression due to the external medium. Cluster mergers can induce shocks or disturbances in the ICM that can compress and lead to detectable emission from such remnant radio galaxies \citep{2001A&A...366...26E,2004rcfg.proc..335K,2019ApJ...870...62P}. Re-acceleration due to turbulence may also play a role as has been suggested in the GREETs (Gently reaccelerated sources) \citep{2017SciA....3E1634D}.


In this paper, we present the detection of steep spectrum antique filaments observed from the multi-frequency GMRT radio maps of the cluster Abell 725 (hereafter A725). We also make use of multi-wavelength archival data from (radio), SDSS (optical) and ROSAT (X-ray) in our detailed study of the cluster. The structure of this paper is as follows. In \S2 we describe the observational and data reduction steps of the GMRT observations. Results of the multi-frequency radio data are described in \S3. Discussion and conclusions of the study are outlined in \S4 and \S5 respectively. We have used the following cosmological parameters: $\rm H_0$ = 70 km\, s$^{-1}$ Mpc$^{-1}$, $\rm \Omega_M$=0.27 \& $\rm \Omega_{\Lambda}$=0.73. These imply a scale of 1.625\,kpc\, arcsec$^{-1}$ at the red-shift z=0.09 of A725.

\begin{figure*}
	\includegraphics[trim={1.0cm 1.5cm 1.5cm 1.5cm}, clip,height=12cm]{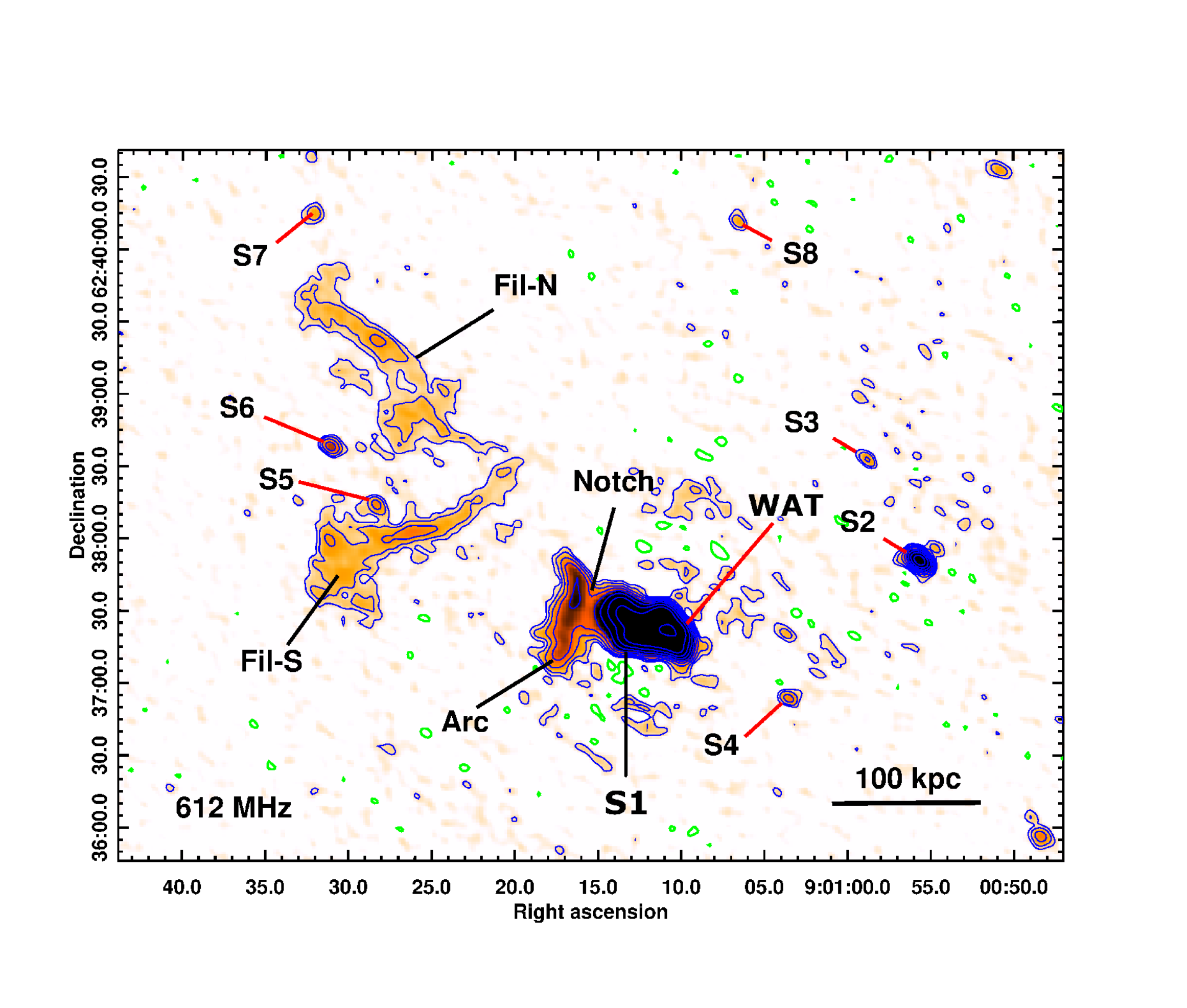}
    \caption{Radio Image of the core of Abell~725 at 612 MHz (GMRT), with rms noise $\sigma$ $\sim$\,$35\mu$Jy beam$^{-1}$, the size of the beam being ($6''\times4''$, $53.1^{\circ}$). The contour levels are $\sigma \times [-3, 3, 6, ...]$. The blue contours are positive and green are negative. Colour within contours is used to indicate different components. The discrete sources are labelled as WAT, S2, ...S8. The source S1 is a radio galaxy, with jets bent towards the East, having features labelled as "Notch" and "Arc". The diffuse emission to the East is labelled as "Fil-N" and "Fil-S", corresponding to the northern and southern filaments.}
    \label{fig:radio_612}
\end{figure*}
\subsection{The poor galaxy cluster A725}
A725 is a poor cluster of galaxies (richness class 0), at a redshift of $z\!=\!0.09$, with an X-ray luminosity $\rm L_{X}(0.1-2.4 \kev)=0.80\times10^{44}$ \lum
and  a velocity dispersion of $\sim$534\kms\, which is consistent with that of a poor galaxy cluster \citep{boschin2008}. The total mass of this cluster within ${\rm R_{500}}$=785\,kpc is estimated to be ${\rm M_{500}}=1.14\times10^{14}\Msun$  \citep{2011A&A...534A.109P}. 
A bright radio source is associated with the brightest cluster galaxy (BCG) of the cluster at the same redshift. In addition,  diffuse radio emission has been detected at the core of the cluster in the WENSS survey, classified as a radio relic \citep{kempnersarazin2001}. 


\begin{figure*}
		\includegraphics[trim={2.1cm 1cm 2.1cm 2.1cm}, clip,height=7cm]{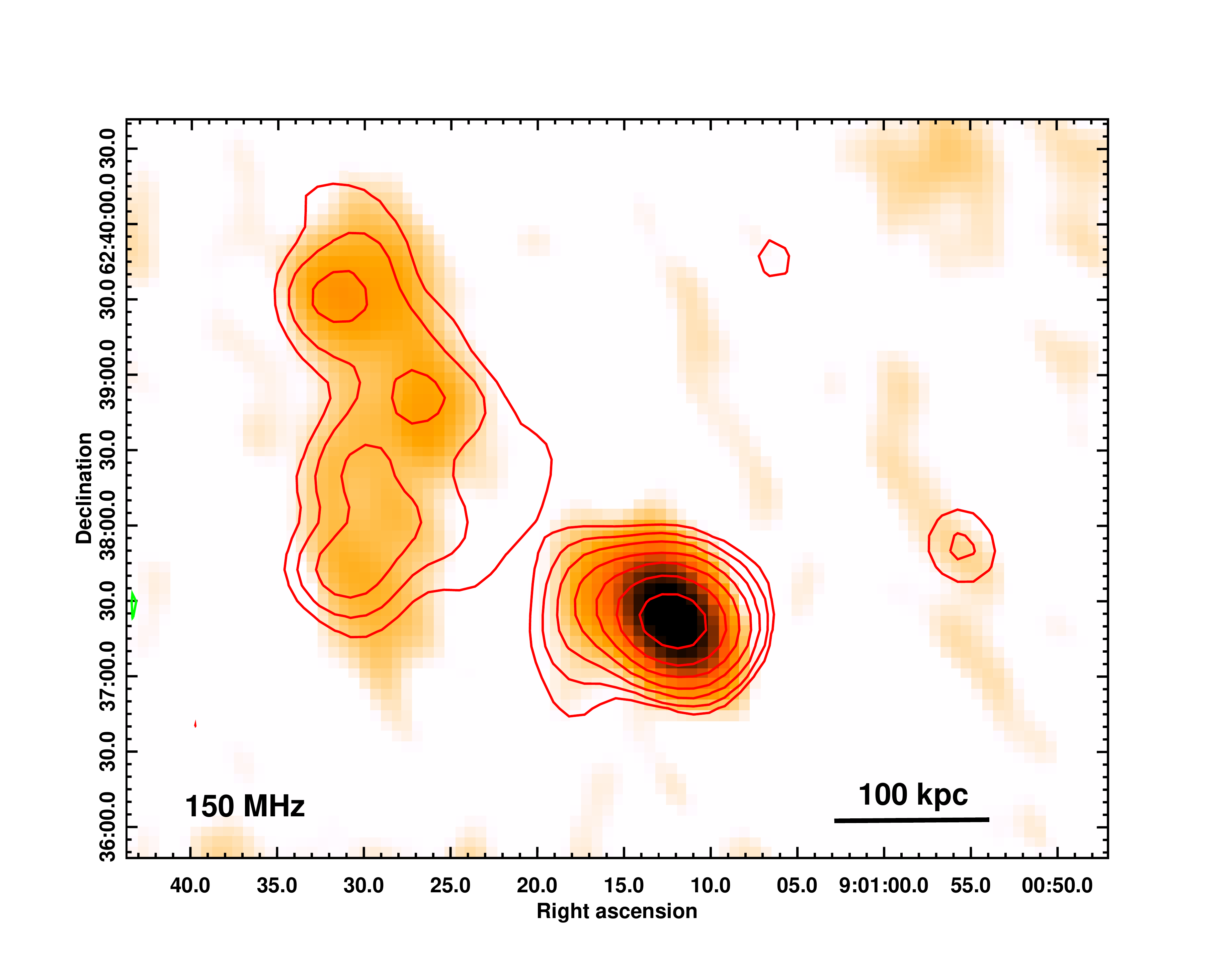}
\includegraphics[trim={2.1cm 2.1cm 2.1cm 2.1cm}, clip,height=7.2cm]{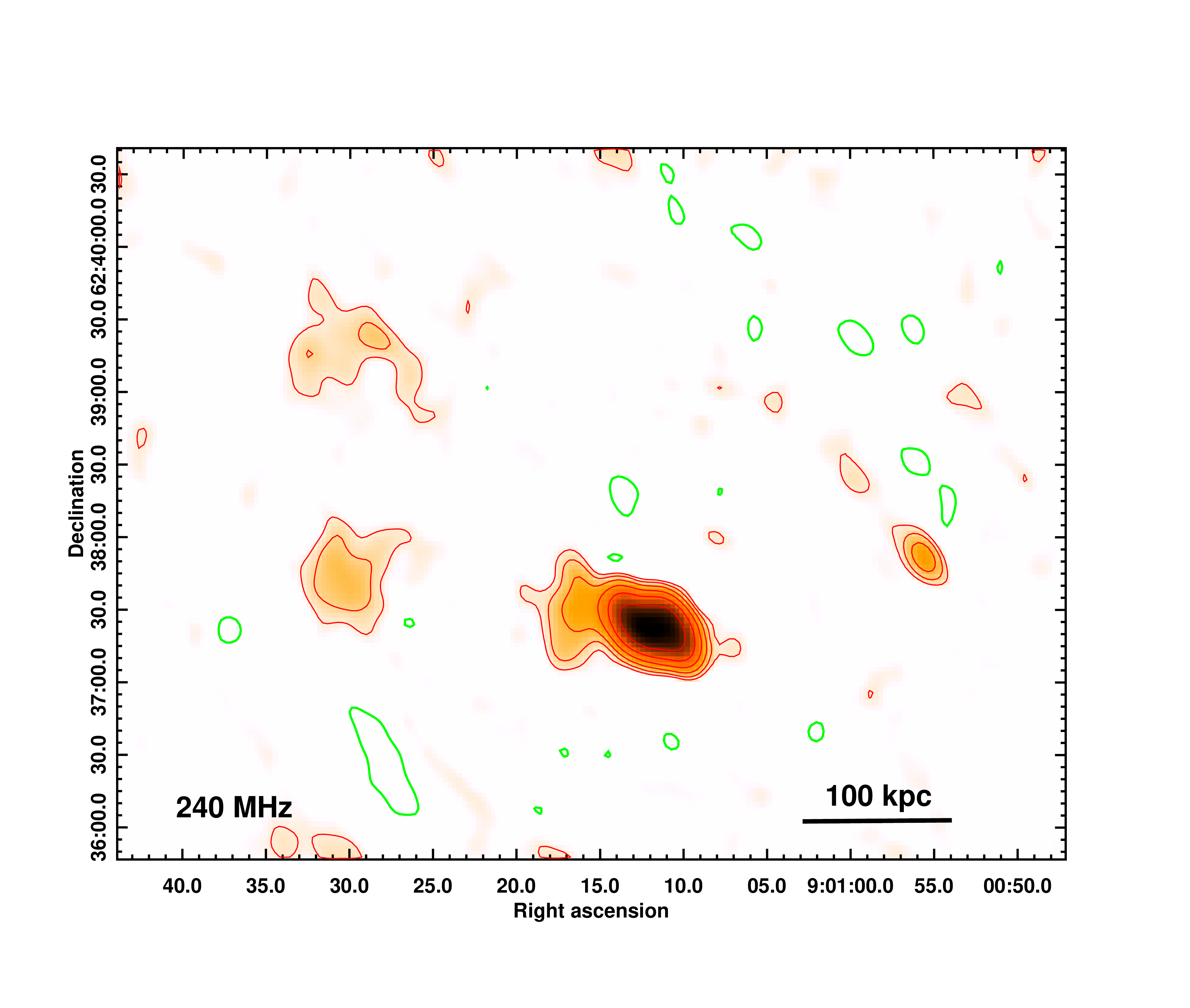}
    \caption{Left: The GMRT 150 MHz image is shown in colour, with contours from the TGSS ADR1 corresponding to 
     $[-5, 5, 10, 20, ...]$ mJy beam$^{-1}$. Right: The GMRT 240 MHz image is shown with contours overlaid. The rms noise in the image is $\sigma=1.1$ mJy beam$^{-1}$ and the contour levels are at $\sigma \times [-3, 3, 6, 12, ...]$. The red contours are positive and green negative in both the panels.}
    \label{fig:gmrt150240}
\end{figure*}

\begin{table*}
 \begin{center}
 \caption{GMRT Observation Details }
 \label{Obs} 
 \begin{tabular}{ccccclcc}
\hline
Central  & Flux  & Phase & Band- & Time on& Observation &rms&beam\\ 
Frequency &\multicolumn{2}{c}{Calibrator} &width & source& Date&&\\ 
(MHz)&&&(MHz)&(h)&& (mJy beam$^{-1}$)& $''\times''$, p. a.\\
\hline
612 &3C286, 3C147 &$1034+564$ &33.3&7 & 03 Jul. 2011&0.035&$6\times4$, $53.1^{\circ}$\\
240 &3C286, 3C147 &$1034+564$ & 6&7 & 03 Jul. 2011 &1.1  &$16.0\times9.9$, $29.9^{\circ}$\\
150 & 3C286  &$0834+555$      & 8&0.8  & 10 Aug. 2007 & 6.0 & $34.3\times20.9$, $20.2^{\circ}$\\ 
\hline
\end{tabular}
 \end{center}
\end{table*}
\section{Observations}
\label{sec:obs} 


\subsection{GMRT observations and data reduction} 
\label{dataan}
A725 was observed with the GMRT \citep{swarup91,swarupgmrt} on two occasions, the proposal codes being 12RKA02 (PI: R. Kale) and $20\_004$ (PI: A. Bonafede), at 150 MHz and in dual-frequency mode (612 and 235 MHz), respectively.
The details of the observations are given in Table ~\ref{Obs}. The 150~MHz observations were carried out with the GMRT Hardware Backend (GHB). The Source Peeling and Atmospheric Modelling (SPAM) \citep{2014ASInC..13..469I} pipeline was used to analyse this dataset. We have also used the TIFR GMRT Sky Survey alternative Data Release 1 (TGSS ADR1, \citet{2017A&A...598A..78I}) 150~MHz image of this field in this study. The flux density scale was set according to \citet[][SH12, hereafter]{2012MNRAS.423L..30S}. The weighting scheme used for imaging is Robust -1. For the purpose of measuring flux we use the TGSS image (observed with the newer GMRT Software Backend or GSB) instead of the 150 MHz observation carried out with GHB.

The dual-frequency observing mode of the GMRT was used to record data in the 612~MHz band in RR polarization and 240~MHz data in LL polarization. The data were recorded in 256 spectral channels across a bandwidth of 33.3 MHz with the GMRT Software Backend \citep{2010ExA....28...25R}. 
The bandwidth used at 240~MHz was 6~MHz. We carried out the data analysis using the CAPTURE pipeline \footnote{\url{https://github.com/ruta-k/uGMRT-pipeline.git}} \citep{2021ExA....51...95K}, which was tailored for the analysis of data recorded in dual-frequency mode. The pipeline is written in python, using tasks in Common Astronomy Software Applications \citep[CASA,][]{2007McMullin} for the data analysis. The standard steps of flagging, calibration, imaging and self-calibration were carried out using the pipeline. The \citet[][PB17]{Perley_2017} scale was used to set the flux density calibration for the dual frequency data. There is agreement between the PB17 and SH12 scales within $2\%$ \citep{Perley_2017}. The images 
were produced with "briggs" weights and robust $= 0$ for the visibilities. The rms noise and beams for the 612 and 240 MHz images are provided in Table~\ref{Obs}. The images have been corrected for the primary beam gain.

\subsection{Other radio data}
In addition to the GMRT data, we have also used high frequency radio data from the Faint Images of the Radio Sky at Twenty-centimeters (FIRST; \citealt{beckerfirst95}), NRAO VLA Sky Survey (NVSS; \citealt{nvss} at 1.4 GHz and Very Large Array Sky Survey (VLASS; \citealt{vlass}) at 3 GHz. These ancillary radio data allowed us to obtain flux measurements at other frequency as well as check and compare morphology at higher resolutions.

For estimating the errors in the flux density measurements, we have used the method of \citealt{klein03}, where flux calibration errors of 3\%, 5\%, 15\% and 20\% for the FIRST (and VLASS\footnote{\url{https://science.nrao.edu/vlass/data-access/vlass-epoch-1-quick-look-users-guide}}), GMRT 612 MHz, GMRT 240 MHz and TGSS (and GMRT 150~MHz), respectively, have been used.

\subsection{Optical and X-ray data}

The earliest systematic multi-band photometric (using the WFC on the Isaac Newton telescope), followed up by spectroscopic (using multi-fibre WYFFOS on WHT) study of A725  \citep{boschin2008} listed 36 member galaxies, with  a mean redshift of $\langle z\rangle=0.0917\pm 0.0003$, corresponding to a line-of-sight velocity of $27~478\pm90$ km s$^{-1}$ with a line-of-sight velocity dispersion of $\sigma_{\rm V}=534_{-97}^{+132}$ km~s$^{-1}$.
The BCG (SDSS J090110.10+623719.6) has a redshift of $\langle z\rangle=0.0900\pm 0.0002$. Further spectral information of the  BCG was obtained from the SDSS database and used for estimating the black hole mass (see \S~\ref{RD} for more details). 

For the X-ray observations, we obtained  broad-band (0.1-2.4$\kev$) pre-processed data from the Roentgen Satellite ({\it ROSAT}) Position Sensitive Proportional Counter (PSPC) {\it ROSAT}All-Skyy Survey Data  (RASS) archives (\url{https://heasarc.gsfc.nasa.gov/}), to study the diffuse X-ray large scale structure. From these RASS images, \citet{kempnersarazin2001} had found that the  X-ray emitting  ICM is elongated in the north-east direction, the same orientation of the claimed radio relic.

\begin{figure*} 
\hspace*{-0.3in}
\includegraphics[scale=0.28]{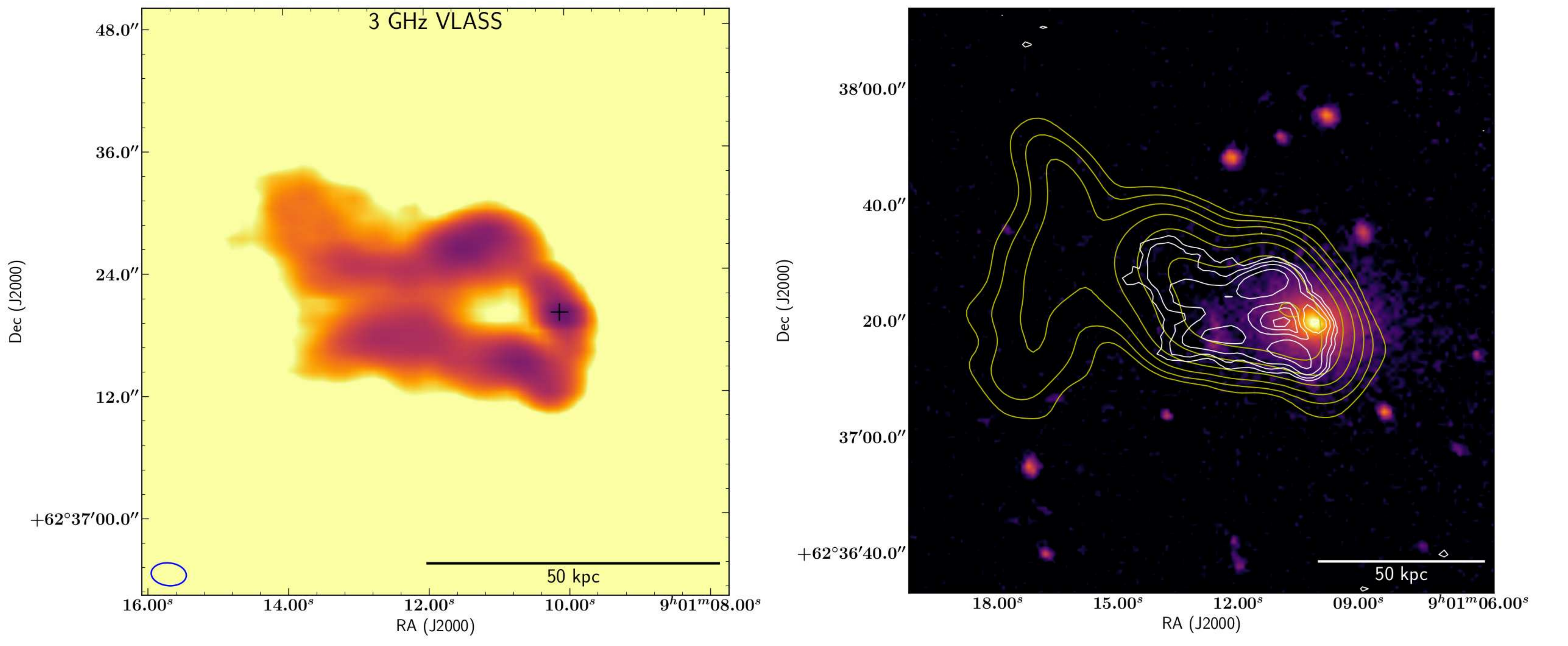}

\caption{Left: 3~GHz VLASS image (3.5\arcsec $\times$ 2.2\arcsec, ~ 84$^{\circ}$ ; rms $\sim$\,0.13 mJy~beam$^{-1}$) of the central BCG of A725 galaxy cluster showing WAT radio galaxy. The cross indicates the location of the BCG. Right: GMRT 612 MHz (yellow) contours and VLASS (white) contours overlaid on  SDSS r band optical image of the BCG and surrounding region.}
\label{fig:cnt_radio}
\end{figure*}

\begin{figure*}
\includegraphics[scale=0.72]{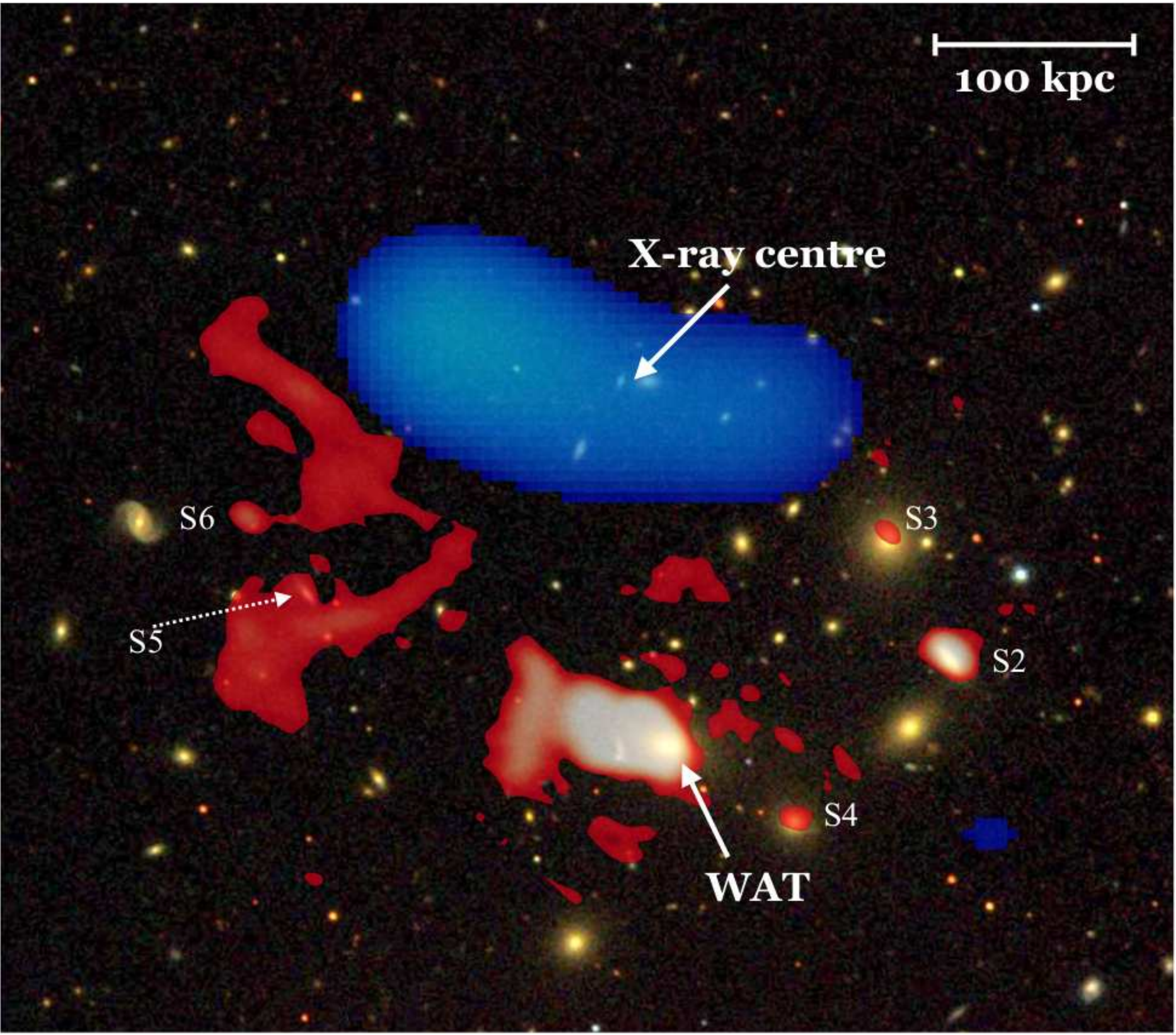}
\caption{Composite image of galaxy cluster Abell~725, with blue representing X-ray emission (ROSAT PSPC), and red denoting radio emission (GMRT 612 MHz). These are overplotted on an optical colour composite ($g$, $r$ and $i$ bands) image from SDSS. }
\label{fig:RGB}
\end{figure*}


\begin{figure*}
	\includegraphics[scale=0.29]{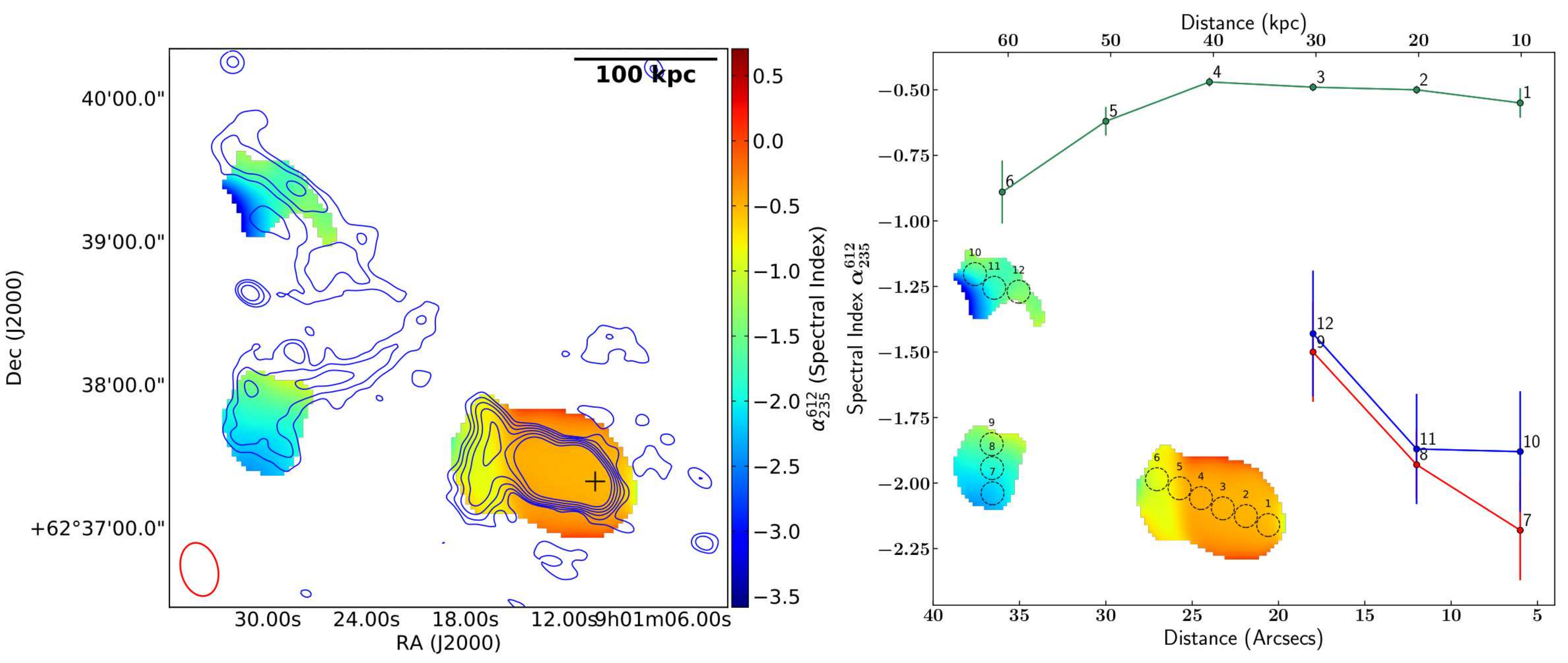}
    \caption{{\it Left A:} Spectral index map of the S1 and the filaments, created using the GMRT 240 and 612 MHz images. The red open ellipse at the bottom left of the image represents the beam of the spectral index map: 22.92\arcsec $\times$ 15.55\arcsec, BPA: 14.52$^{\circ}$. Contours representing the GMRT 612 MHz image are also plotted over the map. The black cross indicates the location of the BCG. {\it Right B:} The spectral index  of the WAT, arc and filaments is plotted with vertical error bars representing 12 circular regions of 10 kpc diameter. }
    \label{fig:Spec_map}
\end{figure*}

\begin{figure}
	\includegraphics[scale=0.067]{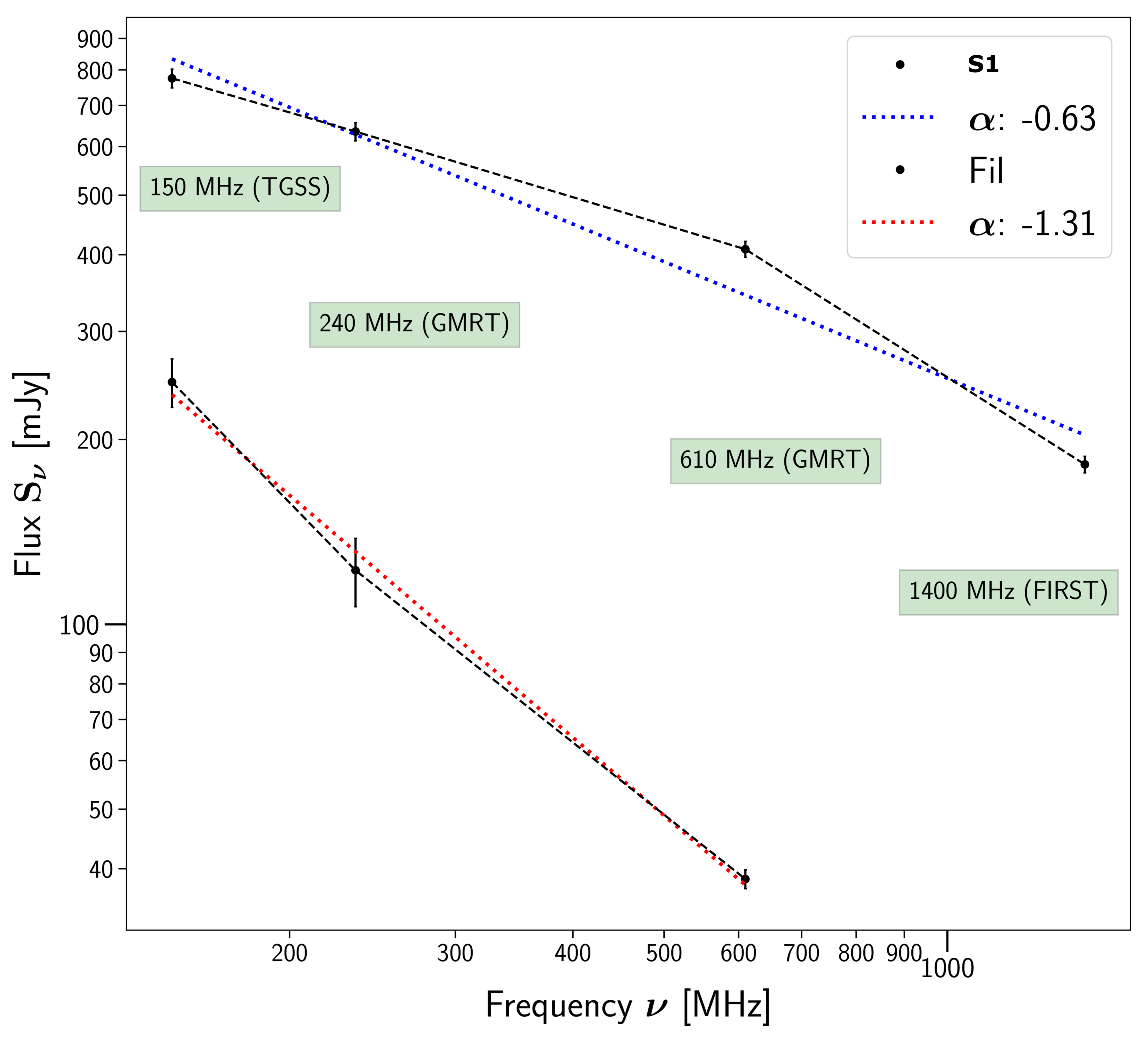}
    \caption{Integrated radio spectra of filaments (combined and considered as a single source) and S1 (upper points) from our GMRT observations are shown here. The FIRST survey 1400~MHz measurement is only used for the S1; the filaments were not detected in this survey. The red and blue dotted lines represent the best fits for respective spectral index values. The flux and spectral index values are presented in Table~\ref{rad-srcs-tab2}. For the S1, the break in the spectrum can be seen at $\sim$\,600 MHz.}
    \label{fig:Spec_plots}
\end{figure}

\begin{figure*}
\hbox
{
\includegraphics[scale=0.09]{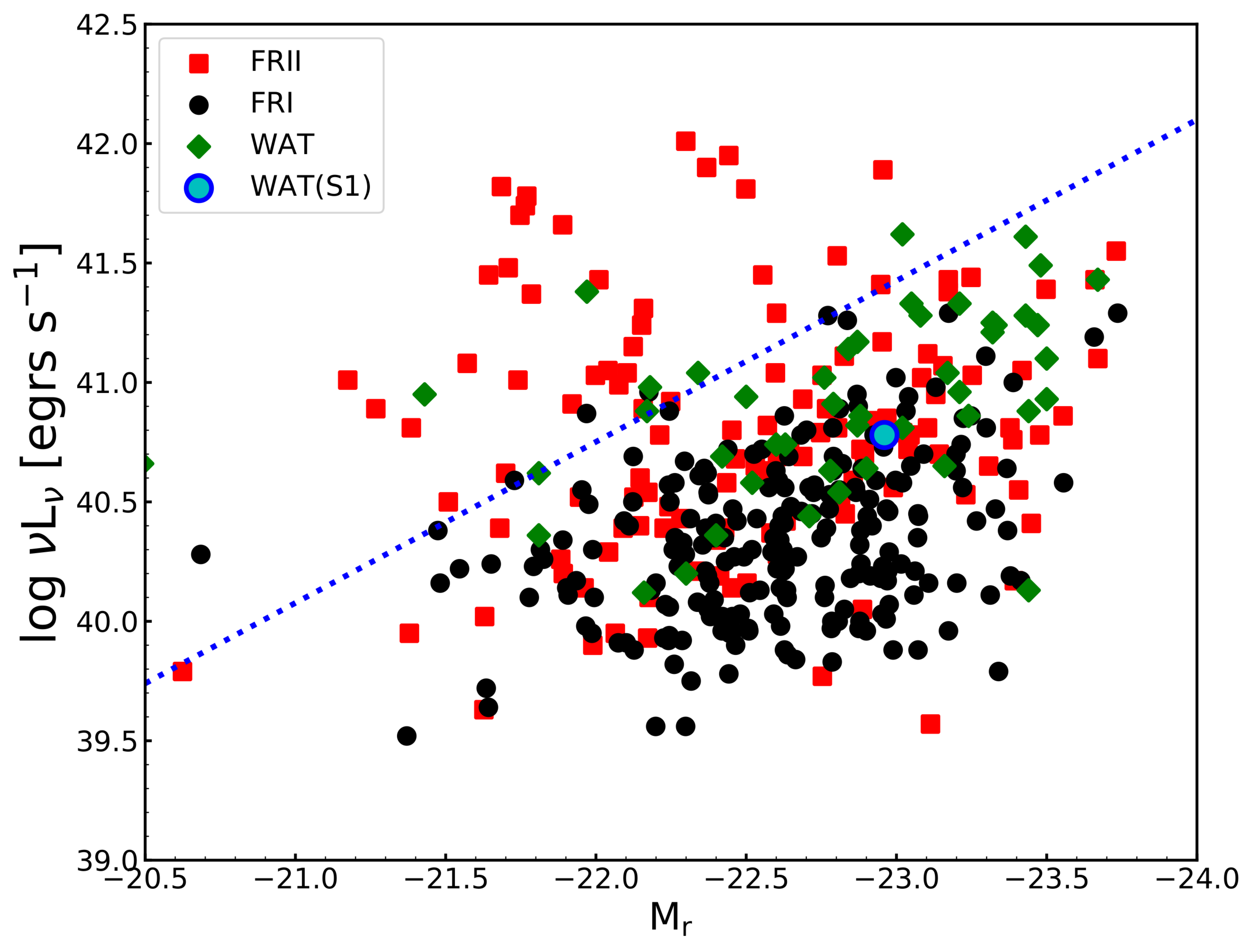}
\includegraphics[scale=0.09]{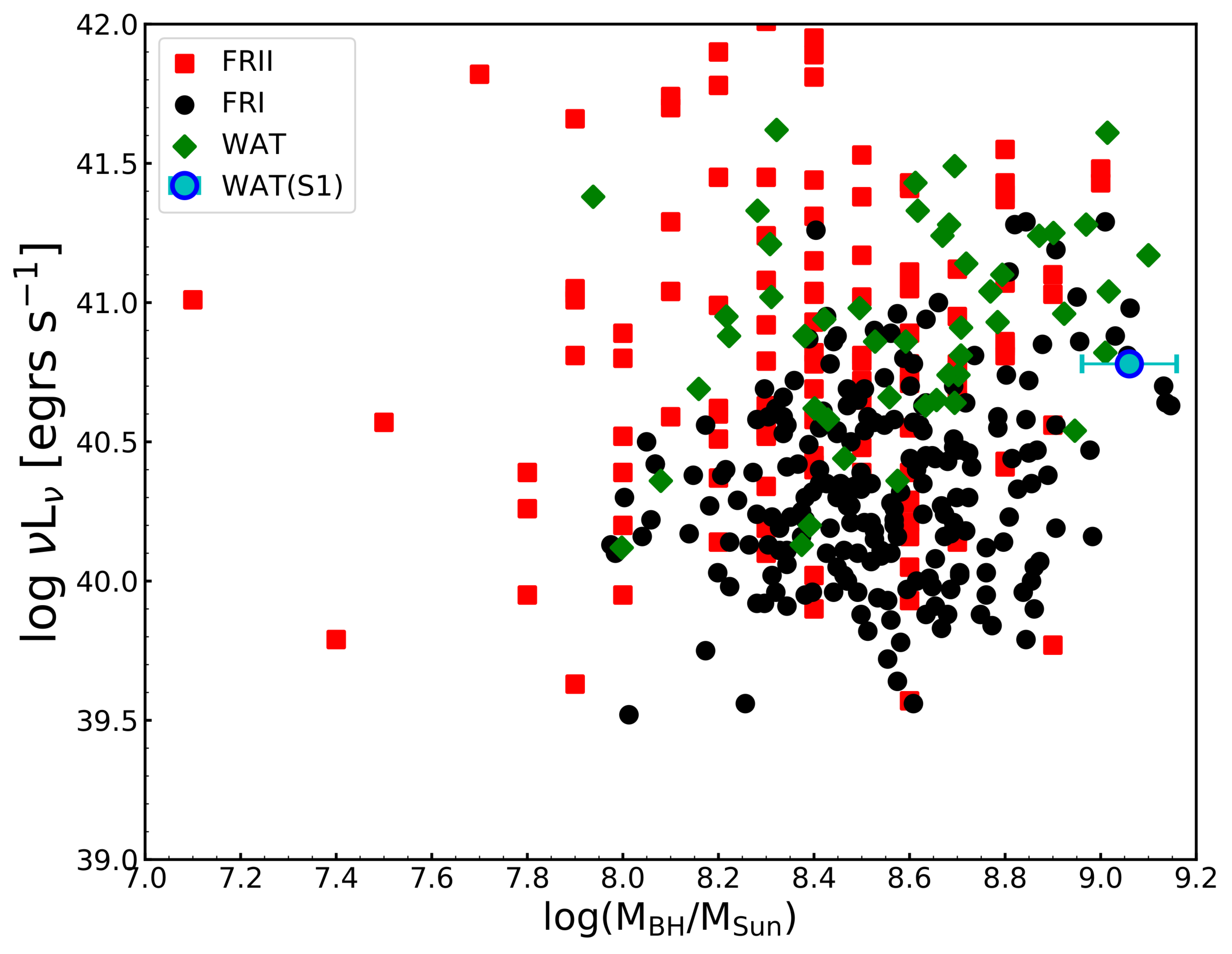}
}
\caption{{\it Left Panel}: Plot of host absolute magnitude ($M_r$) vs. radio luminosity from NVSS at 1.4 GHz, for the WAT, FRICAT, and FRIICAT sources (green diamonds, black dots, and red squares respectively). The dashed blue line shows the separation between FR~I and FR~II \citep{1996AJ....112....9L}.  The WAT source in this paper lies below the boundary between FR~I and FR~II radio galaxies. {\it Right Panel}: The radio luminosity (NVSS at 1.4 GHz) vs. estimated black hole mass for the WAT, FRICAT, and FRIICAT sources (green diamonds, black dots, and red squares, respectively). WAT and FRIs have similar black holes masses, but WATs are more radio-luminous.}
\label{fig:figBH}
\end{figure*}


\section{Results}
\subsection{Radio images}
The GMRT 612 MHz image is shown in Fig.~\ref{fig:radio_612}, where the discrete and diffuse sources detected in this field are labelled for clarity. The source S1 is a  radio galaxy, exhibiting three interesting features, which are labelled as wide-angle-tailed  "WAT", "Notch" and "Arc". 
The diffuse emission to the east seems to correspond to two filaments, marked as the northern and southern filaments, as "Fil-N" and "Fil-S". 

The GMRT 240~MHz image is shown in Fig.~\ref{fig:gmrt150240} (right). The diffuse filaments, radio galaxy with bent jets, the Arc and the Notch are all also detected at this lower frequency, but at a lower resolution. 
Moreover, the radio galaxy and the diffuse filaments are also  detected at 150~MHz in the TGSS (beam = 25\arcsec $\times$ 25\arcsec, rms = 1.9 mJy~beam$^{-1}$) image Fig.~\ref{fig:gmrt150240} (left), though at the resolution of the 150~MHz image (beam $=34'' \times 21''$), the Notch and the Arc features cannot be resolved, however, the 150~MHz image is consistent with the features detected at 612 and 240~MHz.

\subsection{Discrete and diffuse radio sources}

We have detected 8 discrete point sources (including the BCG) in A725 cluster region and their properties are tabulated in  Table~\ref{rad-srcs-tab}.
The source S1, as mentioned in the previous section, is a radio galaxy with jets bent towards the east (Fig.~\ref{fig:cnt_radio}, left). The eastward lobes have a "Notch" and further have a flared morphology that we have labelled as the "Arc". The optical counterpart of the source S1 has been identified as the bright elliptical galaxy SDSS J090110.10+623719.6 ($z=$0.09). 
The BCG coincides with radio core detected in the high resolution ($2\arcsec \times 2\arcsec$) VLASS~3~GHz radio image as seen in Fig.~\ref{fig:cnt_radio}, right. 
We have also examined the 3 GHz VLASS image for correspondence with our other discrete sources.  The source S2 detected at 612 MHz (Fig.~\ref{fig:radio_612}) is also detected at 3~GHz (VLASS) and has a flux density of $12.6\pm0.2$ mJy.
 
The radio point sources present around the BCG have optical counterparts, as can be seen in Fig.~\ref{fig:RGB}. For these, we also measured the radio flux densities using the GMRT 612 and 240 MHz emission maps. The radio and optical properties \citep{boschin2008} of these sources are listed in Table.~\ref{rad-srcs-tab}. 
Sources S7 and S8, which are towards the north, seem to be non-cluster members and will not be considered for discussion further. 

The flux densities of WAT, Arc and the diffuse structures (Filaments)  were measured using the 612 and 240 MHz GMRT images (Table ~\ref{diffuse sources}). We have used the Common Astronomy Software Applications (CASA) \citep{2007McMullin} for measuring the flux densities at the respective frequencies by selecting the regions of radio emission above 3$\sigma$.
The radio powers of the WAT, Arc and filaments associated with A725 were calculated using the relation 
\begin{equation}
P_{\nu_{0}}=4 \pi D_{\rm{L}}^{2} S_{\nu_{0}} (1+z)^{-(1+\alpha)},
\end{equation}
where $D_{\rm{L}}$, $S_{\nu_{0}}$, $z$ and $\alpha$ represent the luminosity distance, flux density at observed frequency $\nu_0$, over the integrated area of the source, redshift and the radio spectral index ($S_{\nu} \propto \nu^{\alpha}$), respectively \citep[e. g.][]{2017A&A...602A...5N}.
The radio power of WAT is the highest among these sources (Table ~\ref{diffuse sources}). We also note that the discrete sources S5 and S6 detected at 612~MHz are located in the region where the diffuse filaments are detected, but they do not show any obvious connection to the filaments.


At 150 MHz, due to the limitation of the resolution, we can separate the filaments and the WAT (Fig.~\ref{fig:gmrt150240}, left) but not the Notch and the Arc. 
In the images with better resolutions (3000, 612 and 240 MHz), we measure the flux density of WAT until the Notch and the flux density of the region after the Notch is considered that of the Arc. Within the filaments, we separately measure the flux densities of the Fil-N and Fil-S.

\begin{table*}
    \begin{center}
    \caption{Discrete radio sources in Abell 725 and their optical identification.} 
    \label{rad-srcs-tab}
    \begin{tabular}{lccccccc}
        \hline
        Source &Right Ascension, Declination &$S_{612\mathrm{MHz}}$  & $S_{240{\mathrm{MHz}}}$  & Optical& R mag & B mag & V  \\
          & hh mm ss, dd mm ss & (mJy)  &(mJy) &Id & & &$\kms$\\
        \hline    
         WAT~~~~ &09 01 09.99, +62 37 20.0  &$383\pm40$ &$592\pm56$  & 24 & 15.12 & 16.41    & 27104$\pm$048 \\
         S2 &09 00 55.68, +62 37 49.5  &$22.7\pm2.2$ & $21.1\pm1.1$ & 19 &15.57 & 17.64 & 29360$\pm$048\\
         S3 &09 00 58.85, +62 38 31.9  &$0.51\pm0.06$ & - & 22 &16.09 & 16.87           & 27739$\pm$049\\
         S4 &09 01 03.51, +62 36 53.9  &$0.62\pm0.07$ & - & 23 &15.62 & 17.67           & 28023$\pm$045\\
         S5 &09 01 28.33, +62 38 13.8  &$0.60\pm0.07$ & - & 27&16.99 & 19.14            & 27616$\pm$103\\        
         S6 &09 01 36.99, +62 38 38.3  & $1.26\pm0.13$ &- & &  &                         &              \\
         S7 &09 01 32.80, +62 40 13.0  & $0.70\pm0.08$  &- & &  &                       &                \\
         S8 &09 01 06.52, +62 40 10.9  & $0.40\pm0.06$  &- & &  &                       &               \\
        
        \hline
    \end{tabular}
    \end{center}
    \end{table*}


\begin{table}
    \begin{center}
    \caption{Radio properties of WAT, arc and the diffuse structures (Filaments) were measured from the 612 and 240 MHz GMRT images.}
    \label{diffuse sources}
    \begin{tabular}{lcccc}
        \hline
        Source &Freq.  & S$_{\nu}$     & $\alpha$~~~~~~~~& $P_{612 \mathrm{MHz}}$\\
               & (MHz)  & (mJy) & & ($10^{24}$ W Hz$^{-1}$)                    \\
        \hline    
            WAT  &612 & $383\pm40$ & $-0.46\pm0.15$& $5.26\pm0.57$ \\
                & 240  & $592\pm56$ &    \\
                \hline    

          Arc & 612 & $23\pm3$ & $-0.80\pm0.30$& $0.32\pm0.04$\\
            & 240 & $49\pm10$& \\
                    \hline    

        Fil-S & 612 & $20\pm3$ & $-1.13\pm0.48$ &$0.29\pm0.04$\\
              & 240 & $58\pm7$  &\\
                      \hline    

         Fil-N & 612&$17\pm3$ &$-1.40\pm0.5$&$0.25\pm0.04$\\
             & 240&$62\pm7$ &\\
 
        \hline
    \end{tabular}
    
     \end{center}
    \end{table}


\begin{table}
    \setlength{\tabcolsep}{4pt}
    \caption{Flux densities of the S1 and filaments separately and together at 150 (TGSS), 240 and 612 MHz from GMRT observations. 
    All the radio maps were convolved to be of the same resolution of 25\arcsec $\times$ 25\arcsec.   
    Figure~\ref{fig:Spec_plots} shows the best-fit spectral index plots of the filaments and the S1. $*$The $\alpha$ for filaments is between 150 to 612 MHz on account of non-detection in the FIRST survey. The total flux at 1400 MHz for filaments and S1 is from NVSS where its unresolved emission is detected.
    }
    \label{rad-srcs-tab2}
    \begin{tabular}{lcccccc}
        \hline
          Region &$S_{150\mathrm{MHz}}$  & $S_{240{\mathrm{MHz}}}$ & $S_{612\mathrm{MHz}}$ & $S_{1400\mathrm{MHz}}$ &  $\alpha_{150}^{1400}$\\
          & (mJy) & (mJy)   & (mJy) & (mJy) &   \\ \hline \hline
         S1 & 775$\pm$27	& 635$\pm$22	& 408$\pm$12	& 182$\pm$6	& -0.63 \\
        \hline
       Filaments(N+S) &248$\pm$23	& 122$\pm$16	& 38$\pm$1.4	& -	 &-1.30$^{*}$ \\
       S1+Filaments &970$\pm$33	& 714$\pm$28	& 441$\pm$13	& 222$\pm$7	  &-0.64 \\
        \hline
    \end{tabular}
    \end{table}

\subsection{Spectral index map and integrated spectra }
In order to make a spectral index map, the UV-coverage at the two frequencies needs to sample the extended source adequately. The GMRT at 612 and 240 MHz can sample sources up to angular scales of $17$ and $44$ arcminutes, respectively. The total extent of the system is less than 4 arcminutes and is consistently sampled at the two frequencies. The final calibrated image of 612 MHz was convolved to  22.92\arcsec $\times$ 15.55\arcsec, BPA: 14.52$^{\circ}$ to match the resolution of 240 MHz.
In Fig.~\ref{fig:Spec_map} (left) and Fig.~\ref{fig:Speciner}, we have shown the spectral index and spectral index error map  of A725 created using the GMRT 240-612 MHz images. The red open ellipse at the bottom left of the image represents the beam: 22.92\arcsec $\times$ 15.55\arcsec, BPA: 14.52$^{\circ}$. The contours correspond to the GMRT 612 MHz image. The black cross mark indicates the location of the BCG. To study the trends of spectral indices in the S1 and the filaments, discrete regions labelled in Fig.~\ref{fig:Spec_map} (right) of 10~kpc diameter each were chosen.
The spectral index as a function of distance is plotted in the same panel. The regions 1--6 show the trend across the S1 from the core to the Arc, in the direction west to east, the regions 7--9 across the Fil-S from south to north and the regions 10--12 across Fil-N from the west to east.  

The flux densities measured at 150 MHz (TGSS) and 1400 MHz are used to create the spectra shown in Fig.~\ref{fig:Spec_plots}, along with the flux densities at 240 and 612 MHz. In order to look for spectral breaks, we constructed the integrated spectra of the radio galaxy and the filaments separately. 
The flux densities measured from the radio images for the S1 and the filaments are reported in Table ~\ref{rad-srcs-tab2}.
The integrated spectra for both the filaments and the S1  are plotted in Fig.~\ref{fig:Spec_plots}.
\subsection{Black hole mass and accretion state}
\label{BM}
The black hole mass of the host galaxy of the BCG is computed using the M$\rm _{BH}$--$\sigma$ relation where $\sigma$ reprsents the effective stellar velocity dispersion \citep{Gultekin}. The value of $\sigma$ of the central host galaxy is 342 $\pm$ 6.5 km s$^{-1}$ which translates into a black hole mass of (1.4 $\pm$ 0.4) $\times$ 10$^{9}$ M$_{\odot}$. Based on the relation from \citet{Heckman2004}, we have estimated the bolometric luminosity (L$_{\rm bol}$) of the central engine to be 2.26 $\times$ 10$^{43}$ erg s$^{-1}$ using the luminosity of the [OIII] emission line. It results into an Eddington ratio ($\lambda_{\rm Edd}$) of 1.24 $\times$ 10$^{-4}$ from the relation  $\lambdaup_{\rm Edd}$ $\equiv$ $\frac{L_{\rm bol}}{L_{\rm Edd}}$, where L$_{\rm Edd}$ is the Eddington luminosity of the black hole. The lower value of $\lambda_{\rm Edd}$ indicates that the BCG is a low-excitation radio galaxy (LERG) with a very low accretion rate \citep{Heckman2014}. In addition, the WISE (Wide-field Infrared Survey Explorer: \citealt{wright10}) mid-IR colour magnitudes (W1 $-$ W2 = 0.065; W2 $-$ W3 = 0.667) indicates the BCG to be a LERG \citep{gurkan14}.
\section{Equipartition magnetic fields}
The minimum energy density (erg~cm$^{-3}$) of a radio source is expressed as \citep{Govoni04},
\begin{equation}
\rm u_{min}= \xi(\alpha,\nu_1,\nu_2) (1+k)^{4/7}  \nu_0^{4\alpha/7} (1+z)^{(12+4\alpha)/7}\times  \\
 I_{0}^{4/7}d^{-4/7},
\label{intnu}
\end{equation}

\noindent
where
$z$ is the redshift of the object, 
$\alpha$ the radio spectral index\footnote{In the formula the spectral index is defined as $S_{\nu}\propto\nu^{-\alpha}$. We have used $\alpha$ with the appropriate sign in the calculations.}, 
$\nu_{1}$ and $\nu_{2}$ are the lower and upper frequency limits in MHz, and the corresponding values are 10 MHz and 10 GHz, respectively. I$_0$ is the surface brightness in mJy~arcsec$^{-2}$
of the object at the break frequency $\nu_0$ (or $\nu_b$ as mentioned in Tab.\ \ref{sa}), 
$d$ the source depth in kpc, and 
$\xi$($\alpha, \nu_1,\nu_2$) are constants whose values are obtained from Table~1 of \citet{Govoni04} for the frequency range 10 MHz--10 GHz. 
This formula accounts for $K-$correction with an assumed value of 1. The parameter $k$ is the ratio of the number density of cosmic-ray protons and electrons.
The magnetic field can be estimated using
\begin{equation}
\rm B_{eq}(cl)= \left({{24\pi}\over{7}} u_{min}\right)^{1/2},
\label{eq:Beq}
\end{equation}
The equipartition energy density ($\rm u_{min}$) and the corresponding magnetic field ($\rm B_{eq}(cl)$) are calculated by employing  classical formalism \citep{Miley}. But the limitation of classical formalism is that it is based on a hardly known parameter i.e. the ratio of the total energy of cosmic-ray protons and electrons. Also, in classical formalism, the spectrum of cosmic-ray electrons is traced over a small range in the observable radio spectrum. As a result, the total energy ratios differ from the energy ratios in the observable range. And in the case of energy losses this difference is more. Therefore, to overcome the limitations of the classical formalism, we also use the revised formalism of \citet{Beck_Krause} to determine magnetic field strength ($\rm B_{eq}$(rev)). It is based on the number density ratio of the particles in the energy range where losses are small and is given as:
\begin{equation}
 \rm B_{eq}(rev) \sim 1.1 \gamma^{\frac{1-2\alpha}{3+\alpha}}_{\mathrm{min}} \times [B_{eq}(cl)]^{\frac{7}{2(3+\alpha)}},
 \label{eq:Beqrev}
\end{equation}
where $\gamma$ is the Lorentz factor and $\gamma_{min}$ is 100.

Using the above relations, we have calculated the magnetic fields (expressed in G) of the S1 and the filaments, as shown in Table~\ref{Tab:Specage}. The spectral indices used in the calculation are 
derived using the integrated spectra shown in Fig.~\ref{fig:Spec_plots}.

\begin{table*}
\centering
\caption{Values of equipartition magnetic field and spectral age estimated for the S1 and the filaments. The break frequency is denoted by ($\rm \nu_b$). The surface brightness I$_0$ is expressed in mJy~arcsec$^{-2}$. Note that the convention $S_{\nu}\propto\nu^{-\alpha}$ has been used for the calculations.}
\label{sa}
\begin{tabular}{cccccccccccc}
\hline
Region & k-value & $\xi$($\alpha, \nu_1,\nu_2$) & I$_0$ & $\alpha$ & $d$ & \textit{u}$_{\rm min}$ &$\rm B_{eq}(cl)$   & $\rm B_{eq}(rev)$ &($\rm \nu_b$)& $\rm \tau_{sp}(cl)$ & $\rm \tau_{sp}(rev)$ \\  
  & & & &  & kpc  & $\times 10^{-11}$ (erg cm$^{-3}$) &$(\rm \mu G $) &( $\rm \mu G) $ & (MHz)  &(Myr)  & (Myr) \\\hline \hline
  
  S1 & 1 & 1.55 $\times$ 10$^{-12}$ & 39.2  & 0.63 & 177 &1.2 & 11.2  & 13.3  & 612 & 46.7 & 37.1  \\
 \hline
Filaments & 1 & 2.79 $\times$ 10$^{-13}$ & 11.1 & 1.31 & 281 &0.34 & 6.1  & 11.3  & 150 & 189.7 & 93.1 \\
      \hline
\end{tabular}
\label{Tab:Specage}
\end{table*}
\subsection{Spectral and Dynamical Ages}
The spectral age of a radio source in Myr can be calculated  using
\begin{equation}
\rm \tau_{sp}=50.3\left[\frac{B^{0.5}}{(B^2+B^2_{IC})}\frac{1}{[(1+z)\rm \nu_{br}]^{0.5}}\right],
\end{equation}
 where $z$ is the redshift, $\rm \nu_{br}$ is the break frequency in GHz \citep{Miley}.
 The value of $\rm B_{IC}$ = 0.318$(1+z)^2$ is magnetic field strength equivalent to cosmic microwave background radiation, and is expressed in nT. 
 
 The magnetic field strength $B$, in $\mu$G, is expressed as $\rm B_{eq}(cl)$ or $\rm B_{eq}(rev)$, depending on 
 whether Eq.~(\ref{eq:Beq}) or Eq.~(\ref{eq:Beqrev}) is used to evaluate the spectral ages. We will denote these as either the classical age $\rm \tau_{sp}(cl)$ or the revised version  $\rm \tau_{sp}(rev)$ of the spectral ages respectively.

From the spectrum of the S1 measured between 150--1400~MHz, a break is expected at  a frequency in the range 612 - 1400~MHz (Fig.~\ref{fig:Spec_plots}). 
Using the lowest possible break frequency of 612 MHz and the magnetic fields described above, we obtain a classical spectral age of 46.7 Myr and a revised spectral age of 37.1 Myr for the S1 (Table~\ref{Tab:Specage}).

For the filaments, the spectrum is a single power-law between 150--612 MHz and the spectral index is --1.31 (Fig.~\ref{fig:Spec_plots}). For a population of relativistic electron population ageing due to synchrotron emission and no new acceleration, the spectral index at injection ($\sim$ -0.7) is expected to steepen by $-0.5$, making the spectral index $\sim-1.2$ \citep{1973ranp.book.....P}. This indicates that the spectral break for the filaments is below 150 MHz. 
If we assume the break frequency to be 150 MHz then we obtain lower limits of classical and revised spectral ages to be 189.7 and 93.1 Myr respectively, for the filaments. The filaments are at least a factor 3 older than the currently active part of the WAT. 
Since we do not have access to the injection spectral index, we note that by using the measured spectral index, we are likely over-estimating the energy of the electrons contributing to the low energy part of the spectrum.

We now calculate the time taken by the BCG to move from the location of the filaments to its current location. The projected distance from the radio core (09:01:06, +62:37:22.0) of the BCG to the Fil-N is 267 kpc and that to the Fil-S is 204 kpc. 
Assuming that the BCG moved from the NE to SW with the same speed as the velocity dispersion of the cluster of 534 km~s$^{-1}$, we obtain the times to traverse the distances from Fil-N and Fil-S to the current location to 
be 489 and 374 Myr respectively. 

The Arc at the eastern end of the WAT is striking in Fig.~\ref{fig:Spec_map} 
as having a steeper spectrum than that of the remaining part of the WAT towards the core. The Notch feature separates the Arc from the remaining WAT 
and may indicate another break in the radio jet activity of the BCG. 
The  distance between the core of the BCG  to the Arc is $\sim$ 85\,kpc, and that between the Arc and the filaments is $92$ (Fil-S) and $204$ (Fil-N) kpc.
The time required to travel 85 kpc by the BCG is 155 Myr. 

The spectral age of the filaments is shorter by a factor $\sim$\, 2.5, compared to the dynamical timescale for the motion of the galaxy. The spectral age can be underestimated due to the overestimation of the magnetic field using the equipartition assumption, the mixing of young and old populations of relativistic electrons and by any processes of re-acceleration that may be at work \citep{2020NewAR..8801539H}. If we were to assume that indeed the dynamical age and spectral age be matched, then the factor by which we over-estimate the magnetic field can be calculated. However, due to continuous evolution in the magnetic field as the lobes expand and possible changes in the motion of the galaxy as it crosses the ICM, the two time-scales can be expected to be discrepant.

Further, in order to understand the effect of ram pressure on radio jets, its orbital motion around the SW sub-cluster, and dynamics of SW cluster, needs high resolutions deeper X-ray telescope observations such as {\it Chandra} and {\it XMM-Newton}. Unfortunately,  high resolution and deeper X-ray data are not available for the A725 cluster in public archives except the {\it ROSAT} shallower sky survey observation. The ROSAT X-ray data  used in this study is inadequate to study the dynamics of S1 and its interplay with the ICM.

\section{Discussion}
\label{RD}

The bent jets associated with the radio galaxy associated with the BCG of the galaxy cluster A725 are resolved in the multi-frequency GMRT radio observations and unveil the steep-spectrum emission further in the direction of the trailing jets towards the north-east (NE). The emission consists of an Arc connected to the trailing jets of the radio galaxy and two filamentary structures further to the east of the Arc at  distances of 204 and 267 kpc from the core.  The filaments do not have obvious peaks indicative of a core - the filaments get completely resolved out at 3 GHz. The sources S5 and S6 are co-located but do not show obvious connections to the filaments.

\subsection{The brightest cluster galaxy of Abell~725}
The optical and X-ray observations indicate that Abell~725 is a merging cluster.
From the distribution of cluster member galaxies in optical images, \cite{boschin2008} found that the cluster is elongated along the east-west (EW) direction and the position of the density peak lies 2~arcmin (195\,kpc) away from the BCG towards the north. They also found a secondary peak in the galaxy distribution NE of the cluster centre, while the cluster itself has an elongated morphology in the NE-SW direction \citep[see Fig.~15;][]{boschin2008}.
The BCG in the cluster is offset from the X-ray emission from the cluster in the ROSAT PSPC image (Fig.~\ref{fig:RGB}). 

The radio morphology of the BCG is consistent with it being a WAT radio galaxy. The jets are bent towards the NE-SW direction, nearly parallel to the elongation of the X-ray emission. The BCG appears to be moving, relative to the cluster from east to west. It is inferred from the jet bending and we considered that BCG may be a part of the NE subcluster before moving to the current position.

The upper jet of WAT radio galaxy seems to be longer and is bent more than that of the lower jet. The projected linear extent of the upper jet measured using high-resolution VLASS image (from the core 09:01:06, +62:37:22.0 to the most outer edges of the jets as identified from the 3 $\sigma$ contours shown in Fig.~\ref{fig:cnt_radio}) is $\sim$ 32 arcsec ($\sim$52 kpc) and that of the lower jet is  $\sim$ 27 arcsec ($\sim$48 kpc).

Since the difference in the radial velocity of the BCG and the mean cluster velocity is 223$\pm$18$\kms$ \citep{boschin2008}, together with the unknown transverse velocity, the net relative velocity would be much higher. BCGs with high velocities with respect to the cluster can occur mainly in merging clusters \citep{2018A&A...609A..78B,2021MNRAS.500..310D}. 
The morphology of the radio jets here is indicative of ram pressure acting on them from the IGM, caused by the high relative velocity, which in this case is likely to be also due to a merger. New observations from GMRT and VLASS suggest that the diffuse emission observed in this system is not a relic but is more likely to be a remnant structure connected with the central AGN activity.

\subsection{Diffuse filaments: past activity of the WAT?}
Some radio galaxies have a duty cycle of repeated activity \citep{saikia09,2020MNRAS.496.1706S}, with a mean lifetime of a few times $10^8$~yr, with a variable period between active periods. The evidence for this can be found in the morphology of the radio emission and in the spectral index distribution. The emission from the older activity is expected to have a  steeper spectrum, due to radiative losses, and to the lack of injection of fresh relativistic plasma. However, sometimes the emission from the previous activity can be detectable while the subsequent activity starts. The morphology is expected to carry signatures of multiple activities in the form of regions of low surface brightness between two activities. If the time of low or no activity is long enough to deplete the relativistic electron population, a discontinuity in the surface brightness may be found.

At the core of the BCG in Abell~725,  the bent jets can be seen to form trailing lobes and at a distance of 50~kpc from the centre, a "Notch" is observed (See Fig.~\ref{fig:radio_612}). The Notch is followed by a structure marked as the "Arc". Beyond the Arc, there is no detected continuum emission till about 92 kpc from the centre of the BCG, where there are two filamentary structures marked in Fig.~\ref{fig:radio_612} as  Fil-S and Fil-N.

The spectral index distribution across the WAT shows an overall flat ($-0.5$) to steep ($-0.9$ in the Arc) trend from the west to east, with the spectral indices in the filaments in the range $-1.5$ to $-2.2$. The constant and flat spectral index values ($\alpha_{612}^{240}= -0.46\pm0.15$) from the  WAT core (region 1) to the end of the tail (region~5) indicate that AGN has been recently active and detected only in the $\sim$\,GHz frequency domain (Fig.~\ref{fig:Spec_map}, right). 
The spectral index value at the position of region~6  (belonging to the Arc) is steeper than that in the core-jet system, and flatter than that in the diffuse filaments, with the Notch feature standing out in the spectral index maps (Fig.~\ref{fig:Spec_map}).

We propose that while the BCG has moved from the NE to SW, along the path joining the filaments, to the current location of the BCG, it has undergone two (possibly three) 
separate bursts of activity, seen in our low-frequency radio observations. The oldest of these bursts of activity is now seen as the steep spectrum filaments, the next younger one being that in the Arc and beyond. It is possible that the currently active core and the jets constitute the third phase of activity.

\subsection{Comparison with other WATs}
\label{CW}
The optical host of the WAT radio galaxy in this cluster is a first-rank bright elliptical galaxy (SDSSJ090110.10+623719), with absolute magnitude $\rm M_r=-22.96$,  its $1.4$~GHz radio power being ${\rm L_{\nu}}=5.03\times 10^{40}$ \lum. We compare this WAT source with those in the WATCAT \citep{2019A&A...626A...8M}, FRICAT \citep{2017A&A...598A..49C}, and FRIICAT \citep{2017A&A...601A..81C} catalogues and shown in Fig.~\ref{fig:figBH} (left panel). In the same figure the absolute magnitudes ($\rm M_r$) of the host galaxies are plotted against the  radio luminosity (NVSS at 1.4 GHz), for the WATCAT, FRICAT and FRIICAT sources (green diamonds, black dots, and red squares respectively). The dashed blue line separates the FR~I and FR~II sources \citep{1996AJ....112....9L}. The WAT source described in this work is shown as a  blue circle and it lies well within the presented distribution (FRI, FRII and WATs). From this figure, it is evident that the absolute magnitude ($\rm M_r$) of the WAT radio source, in the optical R-band, is similar to that of other similar WATs, which are generally associated with the brightest cluster galaxy when they are found  within a cluster. This fact strengthens our argument that the WAT  could be the brightest member of the NE sub-cluster, and now it is in process of merging with the SW sub-cluster.

In the right-hand panel of Fig.~\ref{fig:figBH}, we plot the mass of the supermassive black hole at the heart of the WAT in Abell~725, as a blue circle, along with the black hole masses associated with the other WATs, FRIs, and FRIIs, against the radio luminosity (NVSS) at 1.4~GHz, from the same sample as above, using the same symbols. 
This plot shows that the WAT radio source has a host, whose central supermassive black hole has a mass whose value is among the highest, even though its radio luminosity is close to the median value. The higher black hole mass also points to the formation of WAT in the NE sub-cluster environment, rather than as an isolated galaxy \citep[e.g.][]{2019MNRAS.488L.134L}. Moreover, the median value of radio luminosity and  lower value of Eddington ratio (see \S~\ref{BM}) indicates that the BCG is as a radio-galaxy is in a low-excitation state (LERG).

\section{Conclusions}
\begin{enumerate}
    
    \item  We have found diffuse radio structures in the cluster A~725, comprising of a  WAT radio galaxy, and features resembling  an arc, a notch, and two  filaments, from low-frequency GMRT radio observations at frequencies of 150--612~MHz. We argue that some of these features could be from the past activity of the nucleus of the BCG.\\

     \item The WAT radio galaxy is hosted by the BCG (SDSS J090110.10+623719.6), based on our images at 612 MHz and 3 GHz, using the GMRT and VLASS radio data, respectively. The East-West orientation of the WAT radio galaxy suggests a peculiar motion of the BCG in line with the merger scenario.\\

      \item The  BCG that hosts the WAT appears to be significantly displaced, offset by 2~arcmin (195\,kpc), as projected in the sky plane, from the X-ray centroid  of the cluster, indicating that the cluster is at an early stage of the merger.\\
     
    \item The BCG is at a distance of $\sim$\,85 kpc from the Arc, while the filamentary structures are displaced from the Arc by 92 (Fil-S) and 204 (Fil-N) kpc. We suggest that these structures result from different stages of radio activity along the path of the WAT, and these distances are typical values for such a scenario.\\
    
    \item  The spectral indices of the WAT and Arc are $\alpha_{612}^{240}= -0.46\pm0.15$ and $\alpha_{612}^{240}= -0.80\pm0.30$, respectively. 
    The filaments have two main parts, Fil-N and Fil-S with the spectral indices $\alpha_{612}^{240}= -1.13\pm0.48$ and $\alpha_{612}^{240}= -1.40\pm0.5$, respectively. \\

    \item The spectral index maps show that the WAT has a flat spectrum at the core that steepens gradually up to the Arc, and continues to steepen into the filaments. \\
    
    \item  Based on the morphology of the components, and the progressive steepening of the components from the core of the WAT to the filaments, we propose that this system is a radio galaxy with trailing antique filaments.\\

    \item The radio spectral ageing analysis shows that the spectral ages of the WAT and filaments are 37.1 and 93.1~Myr, respectively. This supports the inference that the filaments and the Arc are features of the older activities of the BCG that have been left behind in the wake of its motion.\\

    \item The radio properties of the  WAT radio galaxy resemble that of typical FRI radio galaxies, as seen in the values of  radio luminosity plotted against $r$-band magnitude. We estimate the mass of the black hole at the core of the WAT, scaling from the  stellar velocity dispersion (342$\pm 6.6\kms$) of the BCG, to be 1.4$\pm 0.4\times10^{9}~\rm \Msun$, which turns out to be rather high compared to that of similar WATs elsewhere. The combination of high BH mass and median  radio luminosity indicates a source of  low Eddington ratio, and that the radio galaxy is a LERG.\\
    
\end{enumerate}

\section*{Acknowledgements}
The authors thank the referee for the comments which have improved the paper.
 MBP gratefully acknowledges the support from the following funding schemes: Science and Engineering Research Board (SERB), New Delhi under the Research Scientist Scheme, 
 Department of Science and Technology (DST), New Delhi under the INSPIRE Faculty  Scheme. 
 RK acknowledges the support of the Department of Atomic Energy, Government of India  
 under project no. 12-R\&D-TFR-5.02-0700. 
 This research has made use of the data from {\it GMRT} Archive.
 We thank the staff of the GMRT that made these observations possible. GMRT is run by the National Centre for Radio Astrophysics of the Tata Institute of Fundamental Research.
 This research has made use of data obtained through the High Energy Astrophysics Science Archive Research Center Online Service, provided by the NASA/Goddard Space Flight Center
This research has made use of  NASA's  Astrophysics Data  System, and of the  NASA/IPAC  Extragalactic Database  (NED) which is operated by the Jet  Propulsion Laboratory, California Institute of Technology, under contract with the National Aeronautics and Space Administration.  Facilities: Chandra (GMRT), SDSS.
This research has made use of the CIRADA cutout service at URL http://cutouts.cirada.ca, operated by the Canadian Initiative for Radio Astronomy Data Analysis (CIRADA). CIRADA is funded by a grant from the Canada Foundation for Innovation 2017 Innovation Fund (Project 35999), as well as by the Provinces of Ontario, British Columbia, Alberta, Manitoba and Quebec, in collaboration with the National Research Council of Canada, the US National Radio Astronomy Observatory and Australia's Commonwealth Scientific and Industrial Research Organisation.

\section*{Data Availability}
The data used in this paper are publicly available at https://naps.ncra.tifr.res.in, https://www.sdss.org and https://skyview.gsfc.nasa.gov. 
The derived data used in the analysis are tabulated in Table.~\ref{Obs}. The code used in reducing the radio observations is publicly available at https://github.com/ruta-k/uGMRT-pipeline.git.

\def\aj{AJ}%
\def\actaa{Acta Astron.}%
\def\araa{ARA\&A}%
\def\apj{ApJ}%
\def\apjl{ApJ}%
\def\apjs{ApJS}%
\def\ao{Appl.~Opt.}%
\def\apss{Ap\&SS}
\def\aap{A\&A}%
\def\aapr{A\&A~Rev.}%
\def\aaps{A\&AS}%
\def\azh{AZh}%
\def\baas{BAAS}%
\def\bac{Bull. astr. Inst. Czechosl.}%
\def\caa{Chinese Astron. Astrophys.}%
\def\cjaa{Chinese J. Astron. Astrophys.}%
\def\icarus{Icarus}%
\def\jcap{J. Cosmology Astropart. Phys.}%
\def\jrasc{JRASC}%
\def\mnras{MNRAS}%
\def\memras{MmRAS}%
\def\na{New A}%
\def\nar{New A Rev.}%
\def\pasa{PASA}%
\def\pra{Phys.~Rev.~A}%
\def\prb{Phys.~Rev.~B}%
\def\prc{Phys.~Rev.~C}%
\def\prd{Phys.~Rev.~D}%
\def\pre{Phys.~Rev.~E}%
\def\prl{Phys.~Rev.~Lett.}%
\def\pasp{PASP}%
\def\pasj{PASJ}%
\def\qjras{QJRAS}%
\def\rmxaa{Rev. Mexicana Astron. Astrofis.}%
\def\skytel{S\&T}%
\def\solphys{Sol.~Phys.}%
\def\sovast{Soviet~Ast.}%
\def\ssr{Space~Sci.~Rev.}%
\def\zap{ZAp}%
\def\nat{Nature}%
\def\iaucirc{IAU~Circ.}%
\def\aplett{Astrophys.~Lett.}%
\def\apspr{Astrophys.~Space~Phys.~Res.}%
\def\bain{Bull.~Astron.~Inst.~Netherlands}%
\def\fcp{Fund.~Cosmic~Phys.}%
\def\gca{Geochim.~Cosmochim.~Acta}%
\def\grl{Geophys.~Res.~Lett.}%
\def\jcp{J.~Chem.~Phys.}%
\def\jgr{J.~Geophys.~Res.}%
\def\jqsrt{J.~Quant.~Spec.~Radiat.~Transf.}%
\def\memsai{Mem.~Soc.~Astron.~Italiana}%
\def\nphysa{Nucl.~Phys.~A}%
\def\physrep{Phys.~Rep.}%
\def\physscr{Phys.~Scr}%
\def\planss{Planet.~Space~Sci.}%
\def\procspie{Proc.~SPIE}%




\bibliographystyle{mnras}
\bibliography{A725} 


 

\appendix
\section{Spectral index error calculations and map}
\label{appendix}

\begin{equation}
\Delta \alpha = \frac{\sqrt{\left(\frac{\Delta S_{\nu_{1}}}{S_{\nu_{1}}}\right)^{2} + \left(\frac{\Delta S_{\nu_{2}}}{S_{\nu_{2}}}\right)^{2}}} {ln \nu_{1} - ln \nu_{2}}
\end{equation}

\begin{figure*}
	\includegraphics[scale=0.30]{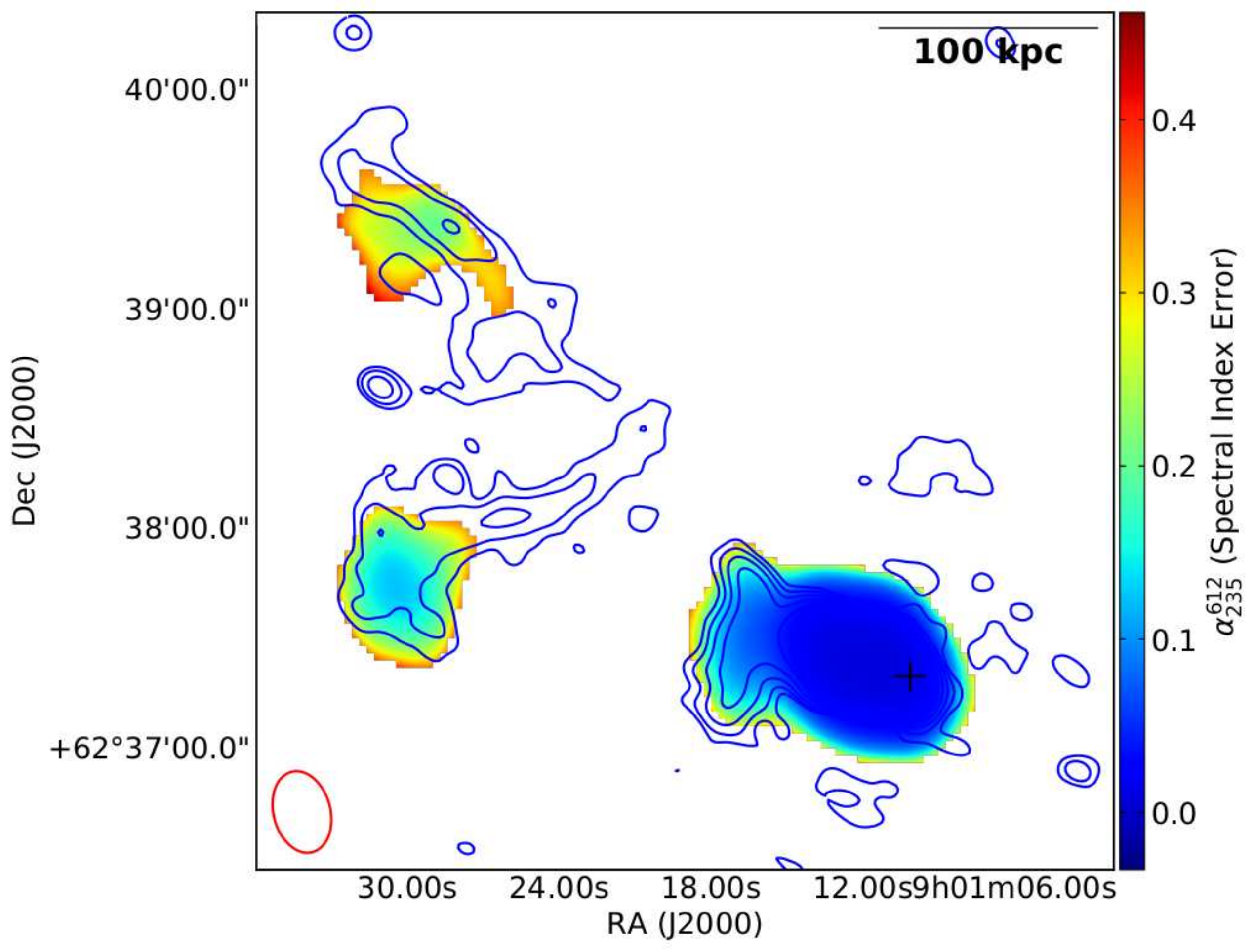}
    \caption{The figure shows the spectral index error map of Fig.\ \ref{fig:Spec_map}.}
    \label{fig:Speciner}
\end{figure*}



\bsp	
\label{lastpage}
\end{document}